\documentclass[prl,twocolumn,showpacs]{revtex4}% Physical Review Letters

\usepackage{graphicx}% Include figure files
\usepackage{dcolumn}% Align table columns on decimal point
\usepackage{bm}% bold math

\begin{document}

%\preprint{APS/123-QED}

%\title{Spectral features of spatially anisotropic triangular antiferromagnets in magnetic fields}
%\title{Spectral properties of anisotropic triangular antiferromagnets in a magnetic field}
\title{Quasiparticles of spatially anisotropic triangular antiferromagnets in a magnetic field}

\author{Masanori Kohno}
\affiliation{WPI Center for Materials Nanoarchitectonics, 
National Institute for Materials Science, Tsukuba 305-0044, Japan}

\date{\today}

\begin{abstract}
The spectral properties of the spin-1/2 Heisenberg antiferromagnet on an anisotropic triangular lattice 
in a magnetic field are investigated using a weak-interchain-coupling approach combined with exact solutions of a chain. 
Dominant modes induced by interchain interactions in a magnetic field behave as quasiparticles which show distinctive features  
such as anomalous incommensurate ordering and high-energy modes. 
In terms of them, various unusual features observed in the anisotropic triangular antiferromagnet Cs$_2$CuCl$_4$ in a magnetic field 
are quantitatively explained in a unified manner. 
\end{abstract}


\maketitle
%{\it introduction.$-$}
In the context of two-dimensional (2D) spin liquids \cite{RVB}, identification of gapless quasiparticles (QPs) 
emerging in the absence of classical long-range orders in 2D frustrated magnets has been 
one of the central subjects in modern condensed-matter physics. Various possibilities of QPs different from conventional magnons, 
including those with fractional quantum numbers called spinons \cite{RVB,ColdeaPRL,ColdeaPRB}, have been 
discussed for 2D spin liquids in the absence or presence \cite{Alicea} of a magnetic field. 
Related to this issue, the anisotropic triangular antiferromagnet Cs$_2$CuCl$_4$ 
has attracted a lot of attention especially because of its spinon-like behavior observed in a zero field \cite{ColdeaPRL,ColdeaPRB}. 
Meanwhile, it shows a host of intriguing features in a magnetic field. Among them are novel incommensurate (IC) ordering \cite{ColdeaPRL} and 
excitation spectra which are hard to understand in terms of magnons in linear spin-wave theory \cite{ColdeaSkw}. 
Nevertheless, the conventional 2D magnon appears above the saturation field $H_s$ \cite{ColdeaModel}. 
Obviously, QPs in the one-dimensional (1D) Heisenberg chain 
in a magnetic field \cite{Karbach_psinon,Karbach_Szz,1DH} cannot consistently explain the above features 
since they show neither IC ordering nor 2D dispersion relations by themselves. 
%Neither conventional magnons nor QPs \cite{Karbach_psinon,Karbach_Szz,1DH} 
%in the one-dimensional (1D) Heisenberg chain can explain the above features consistently, 
%because the 1D QPs do not show IC ordering \cite{ColdeaPRL} 
%or 2D features \cite{ColdeaModel} by themselves. 
\par
To explain the puzzling behavior in a magnetic field, we consider QPs which are defined 
as collective modes induced by interchain exchange processes 
and can be regarded as bound states (BS) or anti-bound states (ABS) of 1D QPs in a magnetic field. 
In zero field, related collective modes \cite{NPhys,Schulz,Essler,Bocquet} may appear 
more or less similar to conventional magnons because of their 
nearly sinusoidal dispersion relations with gapless points close to those of magnons. 
However, the QPs in a magnetic field we consider in this Letter show distinctive features 
which differ from those of conventional magnons \cite{SW} 
and those of 1D QPs \cite{Karbach_psinon,Karbach_Szz,1DH} not only conceptually but also in appearance. 
In particular, we mainly focus our attention on the following features of the QPs: 
(1) multiparticle crossover in a magnetic field, (2) frustration-induced IC ordering whose momentum strongly depends on the magnetization, 
and (3) excitation spectra with anomalous high-energy modes. 
We find, rather remarkably, that various unusual features observed 
in Cs$_2$CuCl$_4$ in a magnetic field \cite{ColdeaPRL,ColdeaSkw,ColdeaModel,CsCuCl_MH} 
are consistently explained as properties of such QPs. 
\par
\begin{figure}
\includegraphics[width=8.7cm]{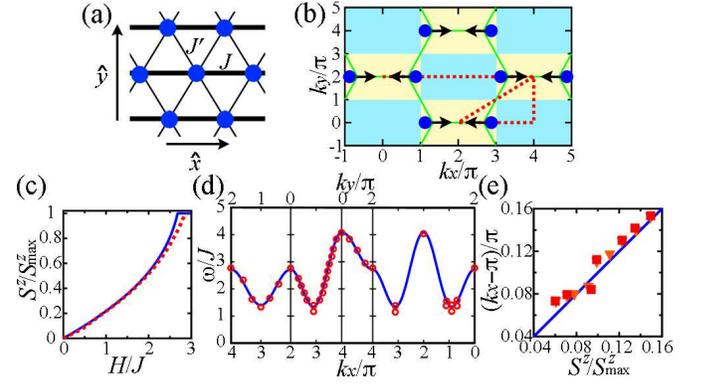}
\caption{(a) Anisotropic triangular lattice. 
(b) Momentum regions of $J'({\bm k})$$>$0 (light blue) and $J'({\bm k})$$<$0 (light yellow) for $J'$$>$0. Solid circles denote 
ordering momenta for $m$=1/16, which shift towards 
the center of each light yellow plaquette as the magnetic field increases as 
indicated by arrows. 
(c)-(e) Comparisons with experimental results on Cs$_2$CuCl$_4$. 
Solid lines are present results. 
(c) Magnetization curve. The dotted line denotes experimental results 
for $H$$\parallel$$c$ in Ref.~\cite{CsCuCl_MH} with $g$-factor  $g$=2.30 \cite{ESR} and $J$=0.374 meV~\cite{ColdeaModel}. 
$S^z_{\rm max}$ denotes $S^z$ at $H_s$. 
(d) Dispersion relation above $H_s$ along dotted lines in (b). 
Symbols are experimental results in Ref.~\cite{ColdeaModel}. 
(e) Momenta of IC ordering. Symbols are 
experimental results at $k_y=0$ for $H$$\parallel$$c$ in Ref.~\cite{ColdeaPRL}, 
replotted using the magnetization curve in~(c).}
\label{fig:lattice}
\end{figure}
%{\it Model and Method.$-$} 
In this Letter, we consider the spin-1/2 antiferromagnetic 
Heisenberg model on an anisotropic triangular lattice 
in magnetic field $H$. The Hamiltonian is defined as 
$$
{\cal H}=\sum_{\bm r}\left(
J{\bm S}_{{\bm r}+{\hat {\bm x}}}
+J'{\bm S}_{{\bm r}+\frac{{\hat {\bm x}}+{\hat {\bm y}}}{2}}
+J'{\bm S}_{{\bm r}+\frac{{\hat {\bm x}}-{\hat {\bm y}}}{2}}\right)
\cdot{\bm S}_{\bm r}-HS^z, 
$$
where ${\bm S}_{\bm r}$ is the spin-1/2 operator at site ${\bm r}$, and 
$S^z\equiv\sum_{\bm r}S^z_{\bm r}$. 
The intra- and interchain couplings and unit vectors are denoted by 
$J$, $J'$, ${\hat {\bm x}}$, and ${\hat {\bm y}}$ as in Fig.~\ref{fig:lattice}(a). 
We assume $J$$>$$|J'|$. In particular, we take $J'/J$=0.34 
corresponding to that of Cs$_2$CuCl$_4$ ($J$=0.374 meV and $J'$=0.128 meV \cite{ColdeaModel}). 
We denote the magnetization per site by $m$. 
The dynamical structure factors (DSFs) are defined as 
$S^{{\bar \alpha}\alpha}({\bm k},\omega)$=$\sum_iM^{{\bar \alpha}\alpha}({\bm k},e_i)\delta$($\omega$$-$$e_i$), where 
$M^{{\bar \alpha}\alpha}({\bm k},e_i)\equiv|\langle{\bm k},e_i|S^\alpha_{\bm k}|{\rm GS}\rangle|^2$ for $\alpha$=$+$, $-$, and $z$. 
Here, $|{\rm GS}\rangle$ and $|{\bm k},e_i\rangle$ each denote the ground state 
and an excited state with excitation energy $e_i$ and momentum ${\bm k}$ 
in the anisotropic-2D system in a magnetic field. 
We also define ${\bar S}({\bm k},\omega)$$\equiv$[$S^{-+}({\bm k},\omega)$+$S^{+-}({\bm k},\omega)$+4$S^{zz}({\bm k},\omega)$]/6. 
\par
We apply two weak-interchain-coupling techniques, 
using exact solutions of the Heisenberg chain \cite{Bethe}. 
One is a random-phase approximation (RPA) \cite{Schulz,Essler,Bocquet}: 
DSFs are obtained from dynamical susceptibilities approximated as 
%$\chi^{{\bar \alpha}\alpha}({\bm k},\omega)\simeq\chi^{{\bar \alpha}\alpha}_{\rm 1D}(k_x,\omega)/[1+J'({\bm k})\chi^{{\bar \alpha}\alpha}_{\rm 1D}(k_x,\omega)/\xi_{\alpha}]$ with $\xi_{z}$=1, $\xi_{\pm}$=2, and 
$\chi^{{\bar \alpha}\alpha}({\bm k},\omega)$$\simeq$$\chi^{{\bar \alpha}\alpha}_{\rm 1D}(k_x,\omega)$/[1+$J'({\bm k})$$\chi^{{\bar \alpha}\alpha}_{\rm 1D}(k_x,\omega)$/$\xi_{\alpha}]$ with $\xi_{z}$=1 and $\xi_{\pm}$=2, where 
the Fourier component of interchain couplings is $J'({\bm k})$=4$J'$$\cos\frac{k_x}{2}$$\cos\frac{k_y}{2}$ for anisotropic triangular lattices, 
and $\chi^{{\bar \alpha}\alpha}_{\rm 1D}(k_x,\omega)$ are calculated from 
DSFs in 1D as 
${\rm Im}\chi^{{\bar \alpha}\alpha}_{\rm 1D}(k_x,\omega)$=$\pi$$S^{{\bar \alpha}\alpha}_{\rm 1D}(k_x,\omega)$ 
for $\omega$$>$0. 
The $S^{{\bar \alpha}\alpha}_{\rm 1D}(k_x,\omega)$ can be calculated~\cite{1DH,Kitanine, Biegel, Caux} using 
rapidities of Bethe-ansatz solutions \cite{Bethe}. 
It is known that spectral properties of the Heisenberg chain in a magnetic field are 
mainly characterized by QPs called psinon ($\psi$) 
and antipsinon ($\psi^*$) \cite{Karbach_psinon,Karbach_Szz,1DH} and 
a QP for a 2-string ($\sigma$) in the Bethe ansatz \cite{1DH,Bethe}. 
We use excitations of up to 2$\psi$2$\psi^*$ \cite{Karbach_psinon,1DH}, 
$O(L^3)$ states of 2-string solutions and those of 3-string solutions \cite{1DH} 
in a chain with length $L$=320 for the RPA. 
These excitations occupy over 80\% of the total spectral weight in $L$=320 
except at very low fields~\cite{1DH}. 
\par
The other technique is the one developed in Ref.~\cite{NPhys}: 
in the restricted Hilbert space spanned by exact eigenstates of chains, 
an effective Hamiltonian is derived as 
\begin{equation}
[{\cal H}^{\alpha}_{\rm eff}({\bm k})]_{i,j}=\epsilon_i(k_x)\delta_{i,j}
+J'({\bm k})A^{\alpha}(k_x,\epsilon_i)A^{\alpha}(k_x,\epsilon_j)/\xi_{\alpha}, 
\label{eq:effHam}
\end{equation}
where $A^{\alpha}(k_x,\epsilon_i)$=$\sqrt{M^{{\bar \alpha}\alpha}_{\rm 1D}(k_x,\epsilon_i)}$. 
It is known that DSFs calculated using the eigenstates of Eq. (\ref{eq:effHam}) are almost the same 
as those of the RPA \cite{NPhys}. 
We discuss physical pictures mainly based on this formulation. 
We use dynamically dominant excitations of $O(L^2)$ states in an $L$=2240 
chain \cite{Karbach_psinon,1DH} and assume 
$S^{{\bar\alpha}\alpha}({\bm k},\omega)$=$M^{{\bar\alpha}\alpha}({\bm k},\omega)D({\bm k},\omega)$ 
with $D({\bm k},e_i)$=2/($e_{i+1}$-$e_{i-1}$)~\cite{Karbach_Szz} for calculations using this technique. 
We neglect instabilities in the ground state for simplicity and show intensities of DSFs in units of 1/$J$. 
\par
This approach is applicable to the full range of momenta, energies, and magnetic fields for 
all directions of spin polarization except small energy scales ($\alt0.1J$) and large $|J'|$ ($\agt0.7J$). 
The wide $J'$ range is due to frustration which prevents 
instability at $k_x$=$\pi$, as will be discussed later. 
The approach reproduces the 2D magnon above $H_s$ and zero-field properties 
obtained in Ref.~\cite{NPhys}. 
\par
\begin{figure}
\includegraphics[width=8.6cm]{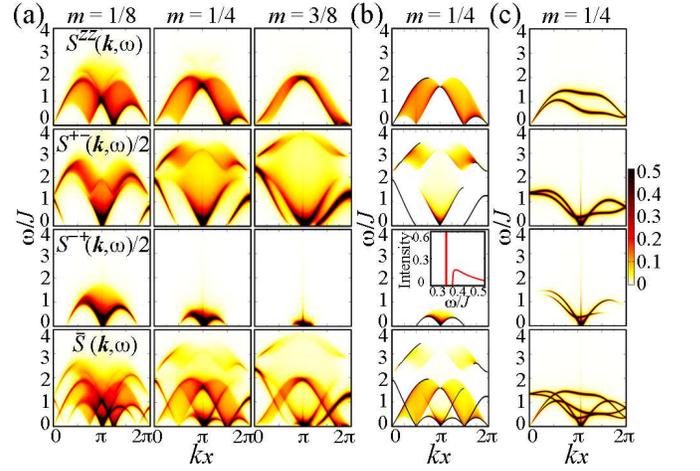}
\caption{$S^{zz}({\bm k},\omega)$, $S^{+-}({\bm k},\omega)$/2, $S^{-+}({\bm k},\omega)$/2, 
and ${\bar S}({\bm k},\omega)$ (from above) at $k_y$=0 for $J'/J$=0.34. 
(a) Results using the RPA at $m$=1/8, 1/4 and 3/8 (from the left), 
broadened in a Lorentzian form with full width at half maximum (FWHM) 0.08$J$. 
(b) Results using the method of Ref. \cite{NPhys} at $m$=1/4. 
Solid lines above (below) continua denote antibound (bound) states 
for 0$\le$$k_x$$<$$\pi$ ($\pi$$<$$k_x$$\le$2$\pi$). 
The inset shows the line shape at $k_x$=1.25$\pi$. 
(c) Single-magnon modes in linear spin-wave theory at $m$=1/4. 
Here, the modes from two cone states, one with the ordering momentum $+{\bm Q}$ and the other with $-{\bm Q}$, 
are shown, broadened in a Lorentzian form with FWHM=0.02$J$.}
\label{fig:Skw}
\end{figure}
As a general property of the approach, spectral weights shift to higher (lower) energies for $J'({\bm k})$$>$0 
($J'({\bm k})$$<$0) [Eq. (\ref{eq:effHam})], and modes with $\delta$-functional line shapes appear above (below) continua 
as poles of $\chi^{{\bar \alpha}\alpha}({\bm k},\omega)$ in the RPA, 
which can be regarded as ABS (BS) of 1D QPs \cite{NPhys,ABS_def}. 
A typical behavior is shown in the inset of Fig. \ref{fig:Skw}(b). 
Although the modes get small intrinsic width due to very small weights around continua, 
the almost $\delta$-functional peaks well represent QPs with integer $S^z$. 
\par
The results for DSFs using the RPA and the method of Ref. \cite{NPhys} are shown in Figs.~\ref{fig:Skw}(a) and \ref{fig:Skw}(b), respectively. 
Strong intensities near upper (lower) edges of continua~for 
$J'({\bm k})$$>$0 [0$\le$$k_x$$<$$\pi$] ($J'({\bm k})$$<$0 [$\pi$$<$$k_x$$\le$2$\pi$]) in Fig.~\ref{fig:Skw}(a) are signatures of ABS (BS), 
whose dispersion relations are asymmetric with respect to $k_x$=$\pi$ due to the sign change of $J'({\bm k})$ at $k_x$=$\pi$. 
Each mode can be identified as follows. In the top row of Fig. \ref{fig:Skw}(a), the 
mode above (below) the continuum in $S^{zz}({\bm k},\omega)$ for $J'({\bm k})$$>$0 ($J'({\bm k})$$<$0) 
can be identified as ABS (BS) of $\psi$ and $\psi^*$ \cite{Karbach_Szz}, 
as indicated by solid lines in the top panel of Fig. \ref{fig:Skw}(b). 
[Dispersion relations as a function of $k_y$ for mod($k_x$,$\pi$)$\ne$0 have a jump at $J'({\bm k})$=0.] 
For $S^{+-}({\bm k},\omega)$, there are three kinds of continua as in the second row of Fig. \ref{fig:Skw}(a). 
The low-energy modes near $k_x$=0 and 2$\pi$ 
can be effectively regarded as QPs originating from the 1$\psi^*$ mode in 1D \cite{Karbach_psinon,1DH,Biegel,Muller}. 
The mode below the low-energy continuum near $k_x$=$\pi$ for $J'({\bm k})$$<$0 can be identified as BS of 
2 $\psi^*$s \cite{1DH}. The mode above (below) the high-energy continua near $\omega$$\agt$2$J$ can be effectively 
regarded as ABS (BS) of $\sigma$ and $\psi$ originating from 2-string solutions 
of the Bethe ansatz \cite{Bethe,1DH}. For $S^{-+}({\bm k},\omega)$, the mode below the continuum for $J'({\bm k})$$<$0 
can be identified as BS of 2 $\psi$s \cite{Karbach_psinon} as in the third row of Fig. \ref{fig:Skw}(a). 
In the following, we discuss distinctive features of these QPs. 
For comparison, behaviors of magnons in linear spin-wave theory \cite{SW,LSWT} are shown in Fig. \ref{fig:Skw}(c). 
\par
\begin{figure}
\includegraphics[width=8.7cm]{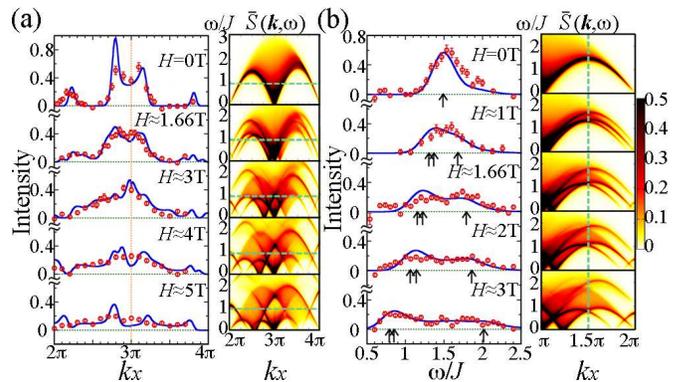}
\caption{Comparisons of line shapes of ${\bar S}({\bm k},\omega)$ with experimental results on Cs$_2$CuCl$_4$ 
at $k_y$=0 for (a) $\omega$=0.35 meV and (b) $k_x$=1.5$\pi$. 
(Left panels) Solid lines are present RPA results of ${\bar S}({\bm k},\omega)$ 
broadened in accordance with the experimental energy resolution~\cite{ColdeaSkw}. 
Arrows in (b) indicate peak positions of $S^{{\bar \alpha}\alpha}({\bm k},\omega)$, 
$\alpha$=$+$, $z$, and $-$ (from the left). 
Symbols denote experimental results for $H$$\parallel$$c$ in Ref.~\cite{ColdeaSkw}. 
Here, the intensities are rescaled after subtracting backgrounds~\cite{ColdeaSkw}, 
and the energies are normalized by $J$=0.374 meV~\cite{ColdeaModel}. 
(Right panels) ${\bar S}({\bm k},\omega)$ using the RPA at $k_y$=0 
corresponding to the left panels, broadened in a Lorentzian form with FWHM=0.08$J$. 
Green dashed lines indicate the scan paths for the left panels.}
\label{fig:cmpexp}
\end{figure}

{\it Multiparticle crossover.$-$}We consider how QP pictures change from 1D spinons interacting through interchain exchange processes 
in zero field~\cite{NPhys} to the 2D magnon above $H_s$ \cite{ColdeaModel}. In zero field, the mode of BS of spinons appears for $J'({\bm k})$$<$0 \cite{NPhys}, 
as in the top right panel of Fig.~\ref{fig:cmpexp}(b). 
Yet considerable spectral weights remain spread in a broad continuum. Hence, 
the latter term in Eq.~(\ref{eq:effHam}), which determines the $k_y$ dependence, has relatively small effects on the dispersion relation. 
Thus, signatures of 1D spinons persist rather strongly in zero field~\cite{NPhys}. 
In a magnetic field, the mode splits into three modes [BS of $\psi$ and $\psi^*$ in $S^{zz}({\bm k},\omega)$, 
BS of $\sigma$ and $\psi$ in $S^{+-}({\bm k},\omega)$, and BS of 2 $\psi$s in $S^{-+}({\bm k},\omega)$], 
as shown in the right panels in Fig.~\ref{fig:cmpexp}(b) and the left panels of Fig.~\ref{fig:Skw}(a) near $k_x$$\simeq$1.5$\pi$. 
As the magnetic field increases, these modes fade away as shown in Fig.~\ref{fig:Skw}(a). 
Instead, the low-energy modes originating from the 1$\psi^*$ mode emerge 
near $k_x$=0 and 2$\pi$ in $S^{+-}({\bm k},\omega)$, 
as in the second row of Fig.~\ref{fig:Skw}(a). 
Using Eq.~(\ref{eq:effHam}) and the 1$\psi^*$ mode in 1D, the dispersion relation is effectively expressed as 
\begin{equation}
\omega({\bm k})=\epsilon_{1\psi^*}(k_x)+J'({\bm k})M^{+-}_{\rm 1D}(k_x,\epsilon_{1\psi^*})/2, 
\label{eq:1antipsinon}
\end{equation}
as shown by solid lines 
near $k_x$=0 and 2$\pi$ at low energies in the second panel of Fig.~\ref{fig:Skw}(b). 
As the magnetic field increases, spectral weights concentrate in the modes, 
increasing $M^{+-}_{\rm 1D}(k_x,\epsilon_{1\psi^*})$ in Eq.~(\ref{eq:1antipsinon}). As a result, 
the modes become dominant and highly dispersing in both $k_x$ and $k_y$ directions like 2D magnons. 
Evidently, as seen above, the dominant modes in low and high fields are different in origin in contrast to magnons in linear spin-wave theory. 
%Thus, unlike magnons in linear spin-wave theory, the dominant modes in high fields are different in origin from those in low fields. 
\par
Noting $H$=$\omega$(${\bm k}$=0)~\cite{resMode,Muller} and using Eq.(\ref{eq:1antipsinon}), we obtain 
\begin{equation}
H=H_{\rm 1D}+4J'm, 
\label{eq:MH}
\end{equation}
where $H_{\rm 1D}$ is the magnetic field in 1D~\cite{Griffiths}. 
This expression is equivalent to that obtained 
in Ref.~\cite{Starykh}. Comparison with experimental results on Cs$_2$CuCl$_4$~\cite{CsCuCl_MH} is shown in Fig.~\ref{fig:lattice}(c). 
Hereafter, we use Eq. (\ref{eq:MH}) to relate $m$ to $H$. 
\par
We confirm that the modes of Eq.~(\ref{eq:1antipsinon}) actually reduce to the 2D magnon above $H_s$. 
By noting that the excitation energy in the Heisenberg chain above $H_s$ 
is given as $\epsilon_{1\psi^*}(k_x)$=$J$($\cos k_x$$-$1)+$H_{\rm 1D}$, 
Eq.~(\ref{eq:1antipsinon}) reduces to the dispersion relation of single-spin-flipped states~\cite{ColdeaModel}: 
$\omega({\bm k})$=$J$($\cos k_x$$-$1)+2$J'$($\cos\frac{k_x}{2}$$\cos\frac{k_y}{2}$$-$1)+$H$ with Eq. (\ref{eq:MH}). 
Small deviations from experimental results on Cs$_2$CuCl$_4$ [Fig.~\ref{fig:lattice}(d)] 
will be due to other effects like Dzyaloshinsky-Moriya (DM) or interlayer interactions in the material~\cite{ColdeaModel}. 
\par
{\it IC ordering.$-$}We consider instabilities at IC momenta expected 
from the behaviors of the BS of $\psi$ and $\psi^*$ in $S^{zz}({\bm k},\omega)$. 
In the Heisenberg chain in a magnetic field, transverse low-energy correlations 
at $k_x$=$\pi$ are always more dominant than longitudinal ones 
at IC momenta \cite{HaldaneXXZ}. 
However, in the presence of frustrated interchain interactions with 
$J'$($k_x$=$\pi$,$k_y$)=0, spectral weights at IC 
momenta for $J'({\bm k})$$<$0 shift to lower energies by interchain interactions 
while those at $k_x$=$\pi$ remain unaffected at least in the present approximation 
[Eq.~(\ref{eq:effHam})]. Thus, instabilities are expected at the momenta where the low-energy 
spectral weight in $S^{zz}({\bm k},\omega)$ becomes the most 
dominant \cite{inc_trans}, 
%i.e. ${\bm k}$=($\pm\pi$(1$+$2$m$)+4$\pi$$l_1$,4$\pi$$l_2$) 
%and ($\pm\pi$(1$-$2$m$)+4$\pi$$l_3$,2$\pi$+4$\pi$$l_4$), 
i.e., ${\bm k}=(\pm\pi(1+2m)+4\pi l_1,4\pi l_2)$ 
and $(\pm\pi(1-2m)+4\pi l_3,2\pi+4\pi l_4),$ 
where $l_i$ ($i$=1$\sim$4) are integers for $J'$$>$0, 
which correspond to the gapless points of the QP for BS of $\psi$ and $\psi^*$. 
The momenta for $m$=1/16 are shown by solid circles in Fig. \ref{fig:lattice}(b). 
Actually, IC ordering whose momentum shifts as 
$k_x$=$\pi$(1$+$2$m$) at $k_y$=0 has been observed 
in Cs$_2$CuCl$_4$ for $H$$\simeq$1$-$2 T$\parallel$$c$ \cite{ColdeaPRL} as 
in Fig. \ref{fig:lattice}(e). Similar behaviors have also been 
observed in Cs$_2$CuBr$_4$ just below the 1/3 
plateau \cite{Ono_JPC}. 
These behaviors have remained unexplained by spin-wave theory \cite{SW}. 
In linear spin-wave theory based on classically stable cone states, ordering occurs at the momentum of 
the gapless point in transverse DSFs [Fig.~\ref{fig:Skw}(c)], which is less dependent on $H$ or $m$ \cite{SW}. 
Note that by higher-order effects of $J'$ for $H$$\simeq$0 \cite{Starykh} and 
by large transverse correlations near $H_s$ \cite{SW,HCB,Starykh}, 
other states can be more stabilized than the IC orders. 
In real materials, ground-state properties are also sensitive 
to DM interactions \cite{SW,HCB,Bocquet,Starykh,CsCuCl_MH}. 
\par
The above argument for the IC ordering will be applicable to 
more general spatially anisotropic frustrated antiferromagnets with 
$J'$($k_x$=$\pi$,$k_y$)=0. (For frustrated 2-leg ladders, we pick up 
2 $k_y$ points.) Thus, the QP picture would be related to the IC ordering in frustrated magnets predicted 
using numerical simulations \cite{Suzuki,eta_inv,Hikihara} and 
field-theoretical analyses \cite{Starykh,Hikihara} and discussed using an RPA for $\omega$=0~\cite{Bocquet}. 
Note that the mechanism of the IC ordering is distinct from that of Ising-like chains~\cite{HaldaneXXZ,KimuraPRL2}. 
\par
{\it Excitation spectra.$-$}Inelastic neutron scattering experiments can probe a quantity 
proportional to ${\bar S}({\bm k},\omega)$. 
We expect behaviors as shown in the fourth row of Fig. \ref{fig:Skw}(a). 
Comparisons with experimental results on Cs$_2$CuCl$_4$~\cite{ColdeaSkw} 
for $\omega$=0.35 meV are shown in Fig. \ref{fig:cmpexp}(a). 
The line shapes can be interpreted in terms of the present QPs. 
For example, the shoulder near $k_x$=2.6$\pi$ in the third row of 
Fig.~\ref{fig:cmpexp}(a) is mainly due to BS of 2 $\psi$s in $S^{-+}({\bm k},\omega)$ 
and BS of 2 $\psi^*$s in $S^{+-}({\bm k},\omega)$. The peak near $k_x$=3$\pi$ comes 
from $\psi\psi^*$ excitations in $S^{zz}({\bm k},\omega)$. 
The dip near $k_x$=3.4$\pi$ is due to the shift of spectral weights near lower edges 
of $\psi\psi^*$ and 2$\psi$ continua to higher energies for $J'({\bm k})$$>$0 [$k_x$$>$3$\pi$]. 
The asymmetry with respect to $k_x$=3$\pi$ is understood as a result of the sign change of 
$J'({\bm k})$ at $k_x$=3$\pi$. 
\par
As in the right panels of Fig.~\ref{fig:cmpexp}(b), the mode of BS 
of spinons in zero field splits into three modes in a magnetic field. 
The left panels show comparisons with experimental results~\cite{ColdeaSkw} 
at ${\bm k}=(1.5\pi,0)$. The low-energy peak is understood as superposition of BS of 2 $\psi$s in $S^{-+}({\bm k},\omega)$ 
and BS of $\psi$ and $\psi^*$ in $S^{zz}({\bm k},\omega)$. 
Interestingly, the high-energy peak turned out to be a signature of BS of $\sigma$ and $\psi$ in $S^{+-}({\bm k},\omega)$ 
originating from 2-string solutions of the Bethe ansatz~\cite{1DH,Bethe}, 
which is not accounted for by the dominant modes in linear spin-wave theory [Fig. \ref{fig:Skw}(c)]. 
\par
In summary, through weak-interchain-coupling analyses using exact solutions of a chain, 
we introduced QPs for anisotropic-2D frustrated Heisenberg antiferromagnets in a magnetic field, 
which can be regarded as BS or ABS of 1D QPs. 
The QPs exhibit distinctive features different 
from those of 1D QPs and those of magnons in linear spin-wave theory. 
Differences with 1D QPs are obvious in view of their $\delta$-functional 
line shapes with integer $S^z$, instability to IC ordering, and 2D features. 
Their statistics may also be different from those of 1D QPs. 
Differences with magnons are the following. 
(1) The QPs show multiparticle crossover in a magnetic field. The dominant QPs in low fields fade away 
in higher fields. Instead, other QPs which reduce to the 2D magnon above $H_s$ emerge in a magnetic field. 
In linear spin-wave theory, magnons in low and high fields are the same in origin. 
(2) The QP in $S^{zz}({\bm k},\omega)$ causes instability to IC ordering whose momentum shifts as 
$k_x$=$\pi$$\pm$2$\pi$$m$ for $J'({\bm k})$$<$0 as in Fig. \ref{fig:lattice}(b). 
In spin-wave theory, it is known that ordering momenta do not shift so significantly as a function of $H$ or $m$ \cite{SW}. 
(3) The high-energy QPs in $S^{+-}({\bm k},\omega)$ come from 2-string solutions of the Bethe ansatz, 
whose behaviors are not explained by linear spin-wave theory. 
\par
These novel features revealed in a magnetic field 
come from the stabilization mechanism of the QPs, which is not based on spontaneous symmetry breaking: 
in contrast to magnons created from classical spin configurations, 
the QPs are induced by interchain exchange processes from liquids of 1D QPs. 
Thus, in analogy to collective modes in Fermi liquids, the QPs may be regarded as 
those in a kind of anisotropic-2D spin liquid or $\psi$, $\psi^*$, and $\sigma$ liquid, distinguished from the magnons. 
The arguments in this Letter can be generalized to weakly coupled antiferromagnetic chains with $J'$($k_x$=$\pi$,$k_y$)=0 
including ladders. 
\par
Their relevance to real materials was confirmed through comparisons 
with experimental results. Various aspects observed 
in Cs$_2$CuCl$_4$ in a magnetic field, such as the 2D dispersion relation, 
IC ordering, line shapes of ${\bar S}({\bm k},\omega)$, 
and the magnetization curve, were explained in a unified manner in terms of the QPs. 
Note that comparisons in this Letter have no adjustable parameters 
except a single normalization factor in Fig.~\ref{fig:cmpexp}. 
It would be interesting in future studies to examine whether signatures of the QPs 
can persist in less frustrated or nearly spatially isotropic 2D antiferromagnets 
in a magnetic field. 
\par
I am grateful to L. Balents, O.A. Starykh, R. Coldea, T. Ono, K. Totsuka, 
A. Tanaka, T. Sakai, T. Hikihara, S. Kimura, M. Takigawa, J. Alicea, and M.P.A. Fisher for 
discussions, helpful comments, and suggestions. This work was supported by 
World Premier International Research Center Initiative on Materials Nanoarchitectonics, 
MEXT, Japan, and KAKENHI 20740206 and 20046015. 
%\bibliography{apssamp}

\begin{references}
\bibitem{ABS_def} ABS are states induced above continua for $J'({\bm k})$$>$0 \cite{NPhys}.
\bibitem{inc_trans} In 1D, $S^{zz}$($k_x$=$\pi$$\pm$2$\pi m,\omega$$\rightarrow$+0) 
is divergent, whereas $S^{\mp\pm}$($k_x$=$\pm$2$\pi m,\omega$$\rightarrow$+0) are vanishingly small.
\end{references}

\end{document}